\documentclass[sigconf]{acmart}

\usepackage{booktabs}
\setcopyright{rightsretained}

\usepackage{balance}  
\usepackage{graphics} 
\usepackage{url}      
\usepackage{fixltx2e}
\usepackage{color}
\usepackage{float}
\usepackage{hyphenat}
\usepackage[flushleft]{threeparttable}
\usepackage[normalem]{ulem}
\usepackage{array}
\usepackage{tabulary}
\usepackage{amsmath,array,graphicx}
\usepackage{kantlipsum}
\usepackage{subfig}
\usepackage{tikz}

\newcolumntype{L}{>{\centering\arraybackslash}m{6cm}}
\newcommand\Tf{\rule{0pt}{2.8ex}}

\acmDOI{10.475/123_4}

\acmISBN{123-4567-24-567/08/06}

\acmConference[WebSci'17]{Web Science Conference}{2017}{Troy, NY} 
\acmYear{2017}
\copyrightyear{2017}
\acmPrice{15.00}

\begin{document}


\title{Using Facebook Ads Audiences for Global Lifestyle Disease Surveillance: Promises and Limitations}

\author{Matheus Ara\'ujo{\small $^{\blacklozenge}$}, Yelena Mejova, Ingmar Weber}
\affiliation{%
  \institution{Qatar Computing Research Institute}
  \city{HBKU, Doha, Qatar}}
\email{maraujo, ymejova,iweber@hbku.edu.qa}
\author{Fabr\'icio Benevenuto}
\affiliation{%
  \institution{Federal University of Minas Gerais $^{\blacklozenge}$}
  \city{UFMG, Belo Horizonte, Brazil}}
\email{fabricio@dcc.ufmg.br}
\begin{abstract}
Every day, millions of users reveal their interests on Facebook, which are then monetized via targeted advertisement marketing campaigns. In this paper, we explore the use of demographically rich Facebook Ads audience estimates for tracking non-communicable diseases around the world. Across 47 countries, we compute the audiences of \emph{marker interests}, and evaluate their potential in tracking health conditions associated with tobacco use, obesity, and diabetes, compared to the performance of \emph{placebo interests}. Despite its huge potential, we find that, for modeling prevalence of health conditions \emph{across} countries, differences in these interest audiences are only weakly indicative of the corresponding prevalence rates. \emph{Within} the countries, however, our approach provides interesting insights on trends of health awareness across demographic groups. Finally, we provide a temporal error analysis to expose the potential pitfalls of using Facebook's Marketing API as a black box.

Please cite the article published at WebSci'17 instead of this arxiv version.


%

\end{abstract}


\maketitle
%

\section{Introduction}

Despite recent media attention on epidemics such as Ebola and Zika, globally, the two most common ``killers'' are ischemic heart disease and stroke\footnote{\url{http://www.who.int/mediacentre/factsheets/fs310/en/}} followed by a set of lung health conditions. Though not as photogenic as Ebola or Zika, lifestyle diseases are in fact far bigger burdens on the global health system than AIDS, influenza and malaria combined. 

Addressing the challenges imposed by \emph{lifestyle diseases} first requires an adequate surveillance system, which not only tracks prevalence rates but also people's awareness of them.  Especially for preventable lifestyle diseases, such awareness is required to induce behavior change.

Recently, encouraging studies linked what people like on Facebook to behavioral aspects and interests in various health conditions~\cite{chunaraetal13pone,gittelmanetal15jmir, gabarron2017impact, brianna2017health,info:doi/10.2196/jmir.6815,araujoetal17icwsm}. These efforts suggest that Facebook data and, particularly, the Facebook advertising system could be a valuable source of behavioral and demographic information, which is key for lifestyle disease surveillance. 

Concretely, Facebook provides its advertisers with access to its users through its targeted advertising platform. Before an ad is launched, and before any cost is incurred, advertisers are provided with an estimate of how many users match the specified criteria, which can include age, gender, sets of interests, specific locations, and much more. 


This type of ``digital census'' opens new avenues for research in public health for several reasons. First, the popularity of Facebook ``Likes'' makes them a promising data source for tracking awareness. Second, the cost of compiling global health statistics such as the Global Burden of Disease database \cite{murraylopez97lancet} could be drastically reduced. Third, the latency of compiling statistics would be reduced, which could make it feasible to measure the impact of policy changes, health campaigns or major global events. Lastly, this approach would allow for an international and cross-lingual analysis by piggy-backing on Facebook's NLP pipeline. Thus, although Facebook Ads is a proprietary system, it is worth investigating its value as a data source for lifestyle disease surveillance. 


In this paper we investigate the feasibility of using Facebook advertising audience estimates for global lifestyle disease surveillance aiming at identifying advantages, disadvantages and limitations of this approach. In particular, we compare these to global disease prevalence across 47 countries, as well as across demographics within the countries. Unlike previous works, we perform quality checks in the form of estimates of ``placebo'' interests, as well as temporal stability of the results. To the best of our knowledge, Facebook ads audience estimates have not been rigorously tested for data quality and stability across countries. Thus, we illustrate the benefits and dangers of using social media as a ``black box'', and provide methodological tools for future research of such social media data aggregators.

\section{Related Work}
\label{sec:related}

In ``Social media in public health'' Kass-Hout \& Alhinnawi assert, ``Social media can provide timely, relevant, and transparent information of public health importance'' \cite{kass-houthend13bmb}, formulating an exciting ongoing research to study health trends via online data. Following this call for action, some researchers focus on information seeking on topics such as abortion \cite{Reis2010MeasuringAbortion} and vaccines \cite{Yom-Tov2014InformationEngine}, others use the data to now-cast diseases \cite{gomide2010dengue}, while others use search logs, most notably using Google Trends, to predict seasonal flu \cite{Ginsberg2009DetectingData}. 

The large errors in flu prediction from Google Flu Trends offered interesting lessons for the use of this kind of data~\cite{lazer2014parable}. While some efforts attempt to fix Google Flu models~\cite{Preis140095}, others have been dedicated to assessing and evaluating the data quality extracted from social media~\cite{kimetal16jmir}.  In this line, we aim at investigating if a new data source can be valuable for health research.

The Facebook advertising API has been studied in terms of determining a relative value of different user demographics and assessing the overall stability of the advertising market \cite{liu2014measurement,saez2014beyond}. A few studies have attempted to link Facebook audience data to behavioral aspects and interests related to health conditions. Gittelman \textit{et al.}\ convert 37 Facebook interest categories to nine factors to use in the modeling of life expectancy \cite{gittelmanetal15jmir}. Although they show an improvement in the statistical models, their approach avoided determining relationships between each individual category with the real-world data. On the other hand, Chunara \textit{et al.}\ explore the relationship of two factors -- interest in television and outdoor activities -- to the obesity rates in metros across the USA and neighborhoods within New York City \cite{chunaraetal13pone}. Both of these studies are confined to the United States, and are limited to one health topic. In contrast, our study expands the set of real-world health indicators we track, as well as the geographic coverage to global proportions. Crucially, we assess the quality of Facebook data by introducing placebo baselines, normalization alternatives, and performing temporal analysis.


\section{Facebook Marketing API}
\label{sec:data}

Facebook, as all large online social networks, relies on advertisement for its revenue. To maximize this revenue it provides advertisers with tools for highly targeted advertising. 
Before launching the ad, and before any cost is incurred, the advertiser is provided with an audience estimate of the number of monthly active users likely to match the criteria.\footnote{The reader can try out an interactive version of the ads tool at \url{https://goo.gl/GNqgUD}.} Facebook provides these so-called ``reach estimates'' via its marketing API.\footnote{\url{https://developers.facebook.com/docs/marketing-api/reachestimate/v2.8}} The available targeting options are truly impressive and, except certain US-specific information related to political leaning or ethnic affinity, are all available in 197 countries worldwide. It is important to highlight that Facebook gathers these estimations based on information from sites other than just facebook.com, as long as those sites have a Facebook Like or share functionality\footnote{http://bit.ly/2dKTJGH}. 

For our analysis, we make use of basic user-provided demographics such as age and gender, and also a set of inferred \emph{interests}. 
According to Facebook, ``Interest may include things people share on their Timelines, apps they use, ads they click, Pages they like and other activities on and off of Facebook and Instagram \footnote{The combined Facebook and Instagram audience were considered for our experiments}. Interests may also factor in demographics such as age, gender and location.''\footnote{\url{https://www.facebook.com/business/help/188888021162119}}. Finally, to automated the data collection, we build a generic python package able to collect data from Facebook Ads \footnote{\url{https://github.com/maraujo/pySocialWatcher}}.

\section{Experimental Setup}

Below we describe how we obtain (i) ``ground truth'' health data, as well as (ii) a set of Facebook-derived features used in our model. 

\textbf{Selection of Health Concerns.} 
To have a well-defined scope, we limit our analysis to three common health concerns: (i) diabetes, (ii) smoking, and (iii) obesity. For all of these, one would expect to find some signal on social media. Furthermore, they cover a mix of one disease (diabetes), one health condition (obesity), and one health-risk behavior (smoking). 
For each, we obtain the prevalence data from the World Health Organization.
For diabetes, we use the \emph{raised fasting blood glucose} statistic for adults (18+)\footnote{\url{http://apps.who.int/gho/data/node.main.A869}};
for smoking, the \emph{adult age-standardized prevalence of smoking of any tobacco product}\footnote{\url{http://apps.who.int/gho/data/node.main.1250}}; and for obesity, the \emph{age-standardized obesity estimates} statistic, where obese is defined as having a body mass index $\geq$ 30\footnote{\url{http://apps.who.int/gho/data/node.main.A900A}}.



\textbf{Selection of Countries. } In order to avoid sparsity issues, we perform our analysis for all countries with more than 5 million active Facebook users, as reported by the API. After removing Taiwan due absence of information in the WHO reports, this resulted in 47 countries, including 12 from Europe, 17 from Asia, 7 from South America, 7 from Africa, 3 from North America, and one in Oceania. 

\textbf{Facebook Marker Interests. } We define a set of \textbf{marker interests} via a bootstrapping approach. We begin with the most obvious interests such as ``Obesity awareness'' for modeling obesity prevalence, and consider other related interests suggested by Facebook Ads explorer\footnote{\url{https://www.facebook.com/ads/manager/creation/creation/}}. We then add to the list those interests which have at least hundreds of thousands in potential audience worldwide. Note that it is not our intention in this paper to define an exhaustive list of Facebook interests related to each health concern, but to start with those with the most plausible direct link to the health concern for this study. The final list of interests can be found in Table~\ref{tbl:interests}. These interests include both those denoting awareness of an illness (such as \emph{lung cancer awareness}), as well as associated behaviors (\emph{Hookah}), and intervention efforts (\emph{smoking cessation}).

\textbf{Placebo Interests. } As latent factors such as socioeconomic status or demographic variables could affect both national health statistics and Facebook interests, correlations themselves might be spurious and not necessarily indicative of any meaningful relationship. To better weed out spurious correlations, we select a set of \textbf{placebo interests}, which are interests on Facebook that should not have an obvious causal link with a given condition, but that might still turn out to be correlated due to latent factors. For this, we use generic interests, including \emph{Facebook}, as well as an OR query of \emph{Reading} OR \emph{Entertainment} OR \emph{Technology}. We also add two health-oriented interests, \emph{Health Care} and \emph{Fitness \& Wellness}, to serve as indicators for whether we are picking up (i) interest specific to a health concern, or (ii) generic interest in health topics. Note that providing a ``placebo'' in such analyses is new, with previous works using Facebook Ads data more as a black box, without determining the merit of using particular marker interests over a baseline \cite{gittelmanetal15jmir,chunaraetal13pone}.

\textbf{Demographic Variables. }
Prior work \cite{anweber15epj} found that the performance for social media based now-casting depends on the subset of users that are being considered. To explore if more fine-grained data could give better models, we also obtained Facebook ad audience estimates for demographic subgroups, in particular for males and females separately, as well as for the age groups 18-24, 25-29, 30-34, 35-39, 40-44, 45-49, 50-54, and 55-59. \footnote{We decided not to include data on minors in our analysis.}


\textbf{User Base Normalization. } At a high level, we are interested in finding whether the fraction of Facebook users in a given country with a particular interest is indicative of a health concern's prevalence in that country. Whether or not a given user has a particular interest is strongly confounded by the time they spent online in general and on Facebook in particular. A random surfer, randomly clicking and liking pages for hours on end, would end up with more inferred health related interests than a user logging in once a month to receive updates from a diabetes forum. In an effort to correct for this, without being able to access the user-level data, we experimented with different normalizers, including \emph{Health Care} and generic topics of \emph{Reading} OR \emph{Entertainment} OR \emph{Technology}. Subsetting to a more specific user set should, in theory, partially remove the confounding effect of different activity levels. Although we found the results to be somewhat different, there was little cohesive trend between data normalized using these specialized populations and one using the Facebook population. Thus, we leave a more detailed study of this issue for future work.

\begin{table}[t]
\caption{Facebook marker interests for tracking tobacco use, obesity, and diabetes, along with placebo interests. Also shown is the estimated worldwide Facebook ad audience.}
\begin{center}
{\footnotesize 
\begin{tabular}{lr}\hline
\textbf{Tobacco Use} & \Tf\\
Smoking & 30,000,000 \\
Tobacco & 20,000,000 \\
Tobacco smoking & 11,000,000 \\
Lung cancer awareness & 6,200,000 \\
Cigarette & 29,000,000 \\
Hookah & 10,000,000 \\
Smoking cessation & 7,500,000 \\
union of all & 77,000,000 \\\hline
\textbf{Obesity} & \Tf\\
Bariatrics & 2,400,000 \\
Obesity awareness & 58,000,000 \\
Plus-size clothing & 29,000,000 \\
Weight loss (Fitness And wellness) & 81,000,000 \\
Dieting & 218,000,000 \\
union of all & 286,000,000 \\\hline
\textbf{Diabetes} & \Tf\\
Gestational diabetes & 1,400,000 \\
Insulin index & 250,000 \\
Insulin resistance awareness & 1,700,000 \\
Diabetes mellitus awareness & 55,000,000 \\
Diabetes mellitus type 1 awareness & 3,200,000 \\
Diabetes mellitus type 2 awareness & 5,300,000 \\
Diabetic diet & 4,200,000 \\
Diabetic hypoglycemia & 280,000 \\
Managing diabetes & 960,000 \\ 
union of all & 60,000,000 \\\hline
\textbf{Placebos/normalizers} & \Tf\\
Facebook & 863,000,000 \\
Reading or Entertainment or Technology & 1,278,000,000 \\
Health Care & 145,000,000 \\
Fitness \& Wellness & 714,000,000 \\\hline
\end{tabular}}
\end{center}
\vspace{-0.5cm}
\label{tbl:interests}
\end{table}

\section{Modeling Prevalence}


\begin{figure*}[!ht]
\begin{centering}
\hspace{-0.5cm}
\subfloat[Placebo]
  {\includegraphics[width=0.25\linewidth]{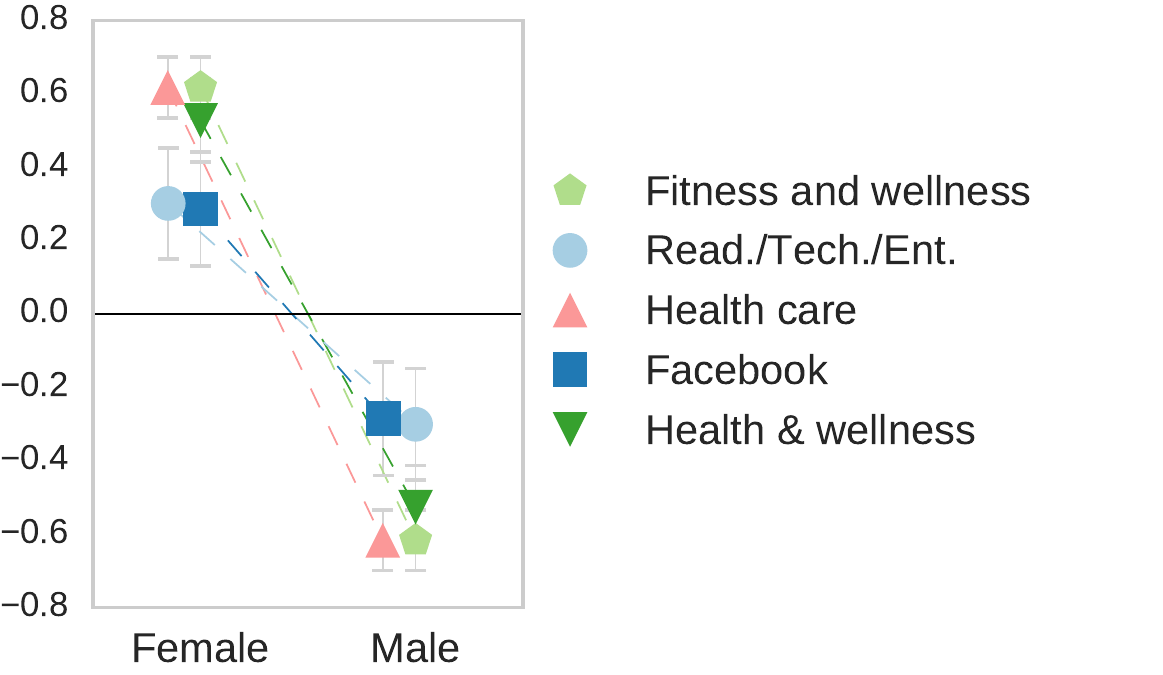}} 
\hspace{-0.2cm}
\subfloat[Tobacco]
  {\includegraphics[width=0.25\linewidth]{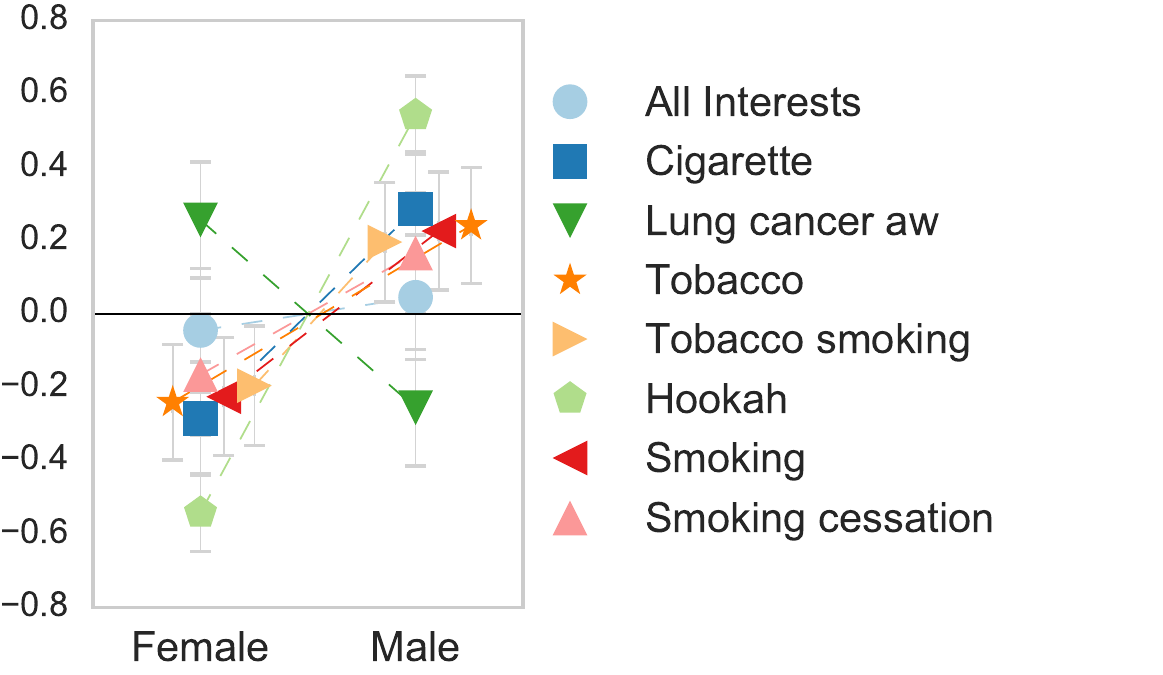}}
\hspace{-0.2cm}
\subfloat[Obesity]
  {\includegraphics[width=0.25\linewidth]{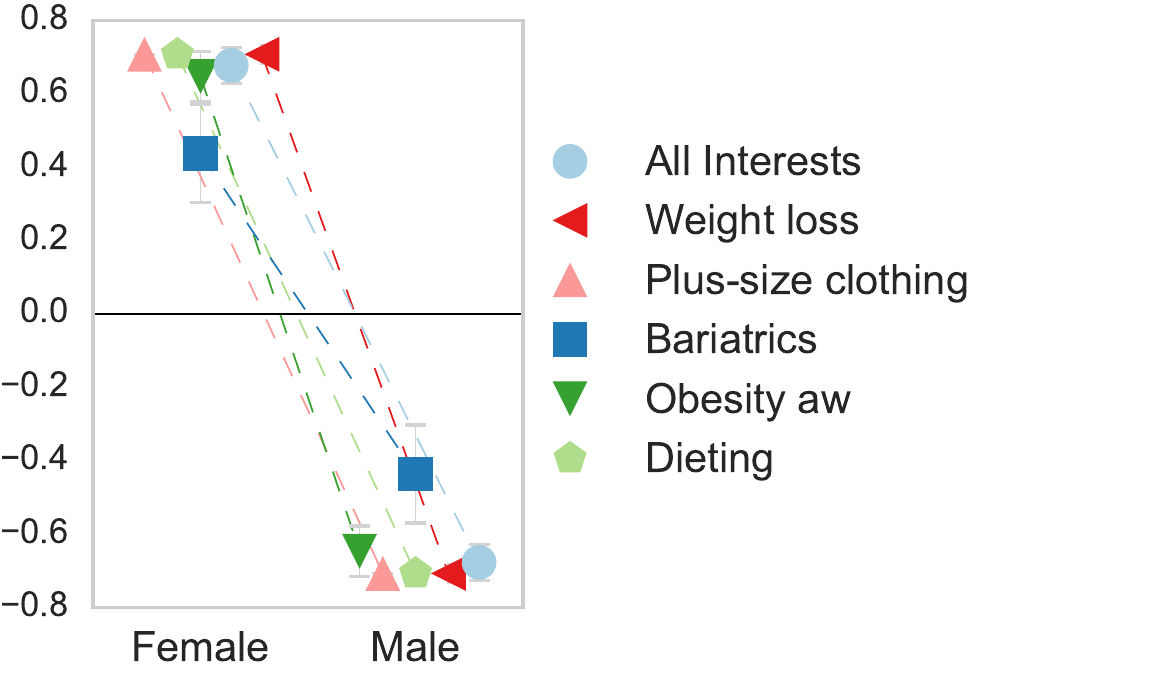}}
\hspace{-0.2cm}
\subfloat[Diabetes]
  {\includegraphics[width=0.25\linewidth]{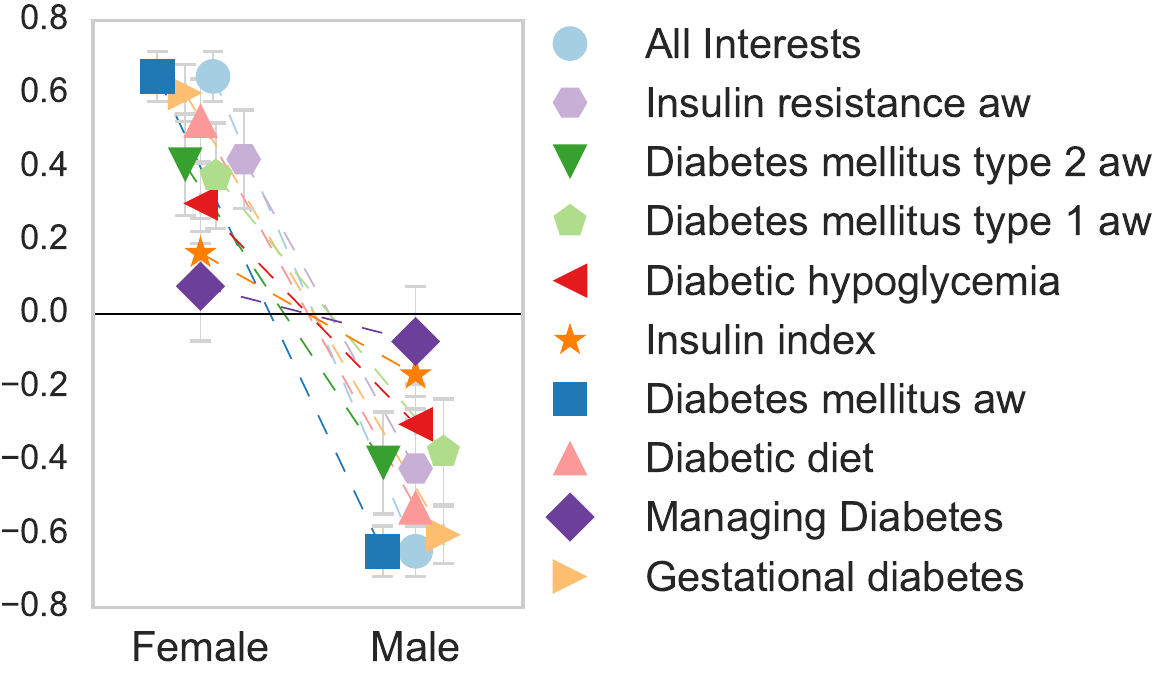}}
\caption{Average $z$-scores per gender of (a) placebo interests, and marker interests for (b) tobacco use, (c) obesity, and (d) diabetes with 90\% confidence intervals.}
\label{figure:zscoregender}
\end{centering}\end{figure*}

\begin{figure}[!ht]
\begin{centering}
\includegraphics[width=0.95\linewidth]{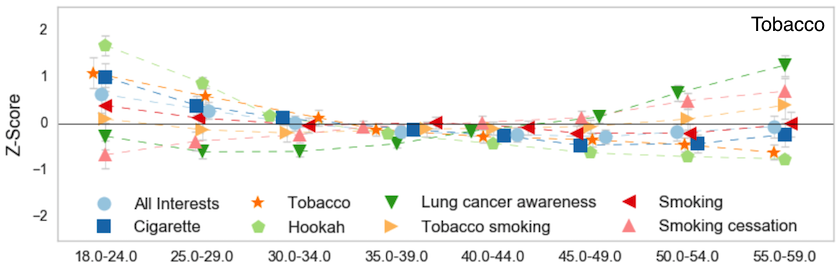}\\\vspace{0.2cm}
\includegraphics[width=0.95\linewidth]{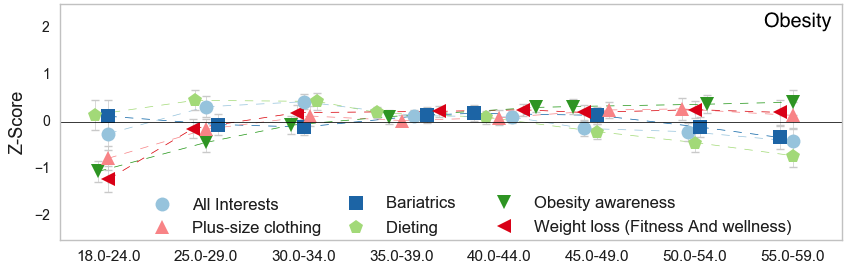}\\\vspace{0.2cm}
\includegraphics[width=0.95\linewidth]{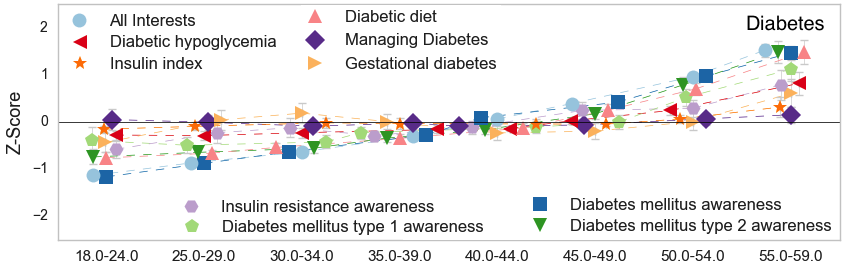}
\caption{Average $z$-scores aggregated for age groups of marker interests for tobacco use, obesity, and diabetes with 90\% confidence intervals.}
\label{figure:zscoreage}
\vspace{-0.5cm}
\end{centering}\end{figure}

Using the above design, we collected the data on October 19, 2016. Overall, 583,200 requests to the Facebook Marketing API were required, and the collected data is available for the community.\footnote{We will make it available upon publication of this article}.

We begin by computing Pearson correlation $r$ of normalized Facebook audience estimates with the corresponding WHO data for each health concern. We find that just two -- the interest \emph{Dieting} with ($r=0.314$) and the OR query \emph{``Obesity awareness OR Plus-size clothing OR Weight loss OR Dieting OR Bariatrics"} ($r=0.344$) -- have a significant correlation to WHO obesity reports. However, we find the placebo interests too often outperform the targeted marker interests. The strongest significant correlations we find are between \emph{Fitness \& Wellness} and obesity ($r=0.590$), and \emph{Health care} and \emph{Facebook} and diabetes ($r=0.313$ and $r=0.302$, respectively). Note that with Bonferroni correction for multiple hypothesis testing, even these results may be questioned. The results do not improve when we break down the data by the age and gender group (plots omitted for brevity). Finally, we would like to emphasize the performance of the baseline, generic health interests, often ``outperforming'' the targeted marker interests. Even the generic placebo interests like \emph{Facebook} and \emph{Read./Tech./Ent.} are often near the top of the list of predictors. 

We then turn to comparisons within each country, across gender and age groups. Concretely, we are interested in whether there are certain generalizable age or gender related trends concerning the variation of interest levels in health topics within a country. To obtain such relative trends, we first compute the absolute interest levels for the relevant demographic group. Then we normalize these statistics within each country by computing the $z$-score (subtracting the mean and dividing by standard deviation across the demographic groups).

Figure~\ref{figure:zscoregender} shows the average $z$-scores of (a) placebo interests, and marker interests for (b) tobacco, (c) obesity, and (d) diabetes, broken down by gender. We find the placebo interests to be expressed more by women than men, especially those dealing with fitness and health. One potential explanation for this is that women are generally more active on social media than men, in particular on mobile devices\footnote{\url{http://www.ukom.uk.net/news/women-driving-mobile-internet-time}}. Yet we also find that women show more interest in topics around obesity and diabetes, though it is expected that, for example, \emph{Gestational diabetes} would be a more popular topic for women than men. We see the opposite for the topics associated with tobacco, with most topics being more favored by men. The only exception is \emph{Lung cancer awareness}, which is, again, a topic associated with health. These findings echo the global statistics of tobacco use, with WHO estimating that ``about 40\% of men smoke as compared with nearly 9\% of women''\footnote{\url{http://www.who.int/gender/documents/10facts_gender_tobacco_en.pdf}}.

Figure~\ref{figure:zscoreage} shows the similar average $z$-scores for various age groups. Examining tobacco-related interest, we find the younger users being more interested in \emph{hookah}, \emph{tobacco}, and \emph{cigarettes}, whereas the older users in \emph{lung cancer awareness} and \emph{smoking cessation}. The users in the age between 30 and 45 do not show preference for one or the other interest. Comparing these figures to the WHO global report on trends in prevalence of tobacco smoking report\footnote{\url{http://apps.who.int/iris/bitstream/10665/156262/1/9789241564922_eng.pdf?ua=1}}, we observe that the reported evolution of tobacco use by age-groups follows the z-score trend of related interests from Facebook Ads. 

Thus, we find that within-country statistics are more statistically separable, which can then be compared to a corresponding ``gold standard''. Next we focus on error analysis.


\section{Error Analysis}

The greatest advantage of Facebook advertising platform is also its weakness -- the extensive natural language processing (NLP) pipeline and an aggregator spanning over a billion of users and many Facebook properties. To us, this black box provides an aggregation on planetary scale, but it also presents a murky opacity. 

As mentioned in Related Work, a similar difficulty faced researchers using Google Trends, which aggregates search volume worldwide, and which potentially can serve as a signal of Google users' interests. A major advantage of Google Trends is its temporal nature, allowing researchers a dimension in which to experiment with various time lags \cite{Preis140095}. Similarly, we re-collect the data for the 10 countries with largest Facebook audiences nearly five months later, on 12 March, 2017, to examine its stability. Then, we compute the correlations between the previous and new audience estimates.

When we consider the audience sizes in each country without constraint by interest, sliced across the age (8) or gender (2) demographic groups, we find the data quite stable, with Spearman rank correlation being 1 for most (in two cases 0.99), and Pearson on average at 0.999. Although there were changes, on average the estimates increased by 14\% (stdev of 8.9\%), which may be due to Facebook's growing usage or increased user tracking reach. 

Once we include the distinction by interest (both marker and placebo ones), the correlation between the data from October 2016 and March 2017 degrades. Across each country and interest ($N$ = 10 countries * 29 interests = 290), we run two stability checks, first for both genders combined we compare the relative audiences for different age groups between the two time samples; second, for all combined ages we compare the audiences for the two genders between the two time samples. 

\begin{figure}[h]
\begin{centering}
\includegraphics[width=0.48\linewidth]{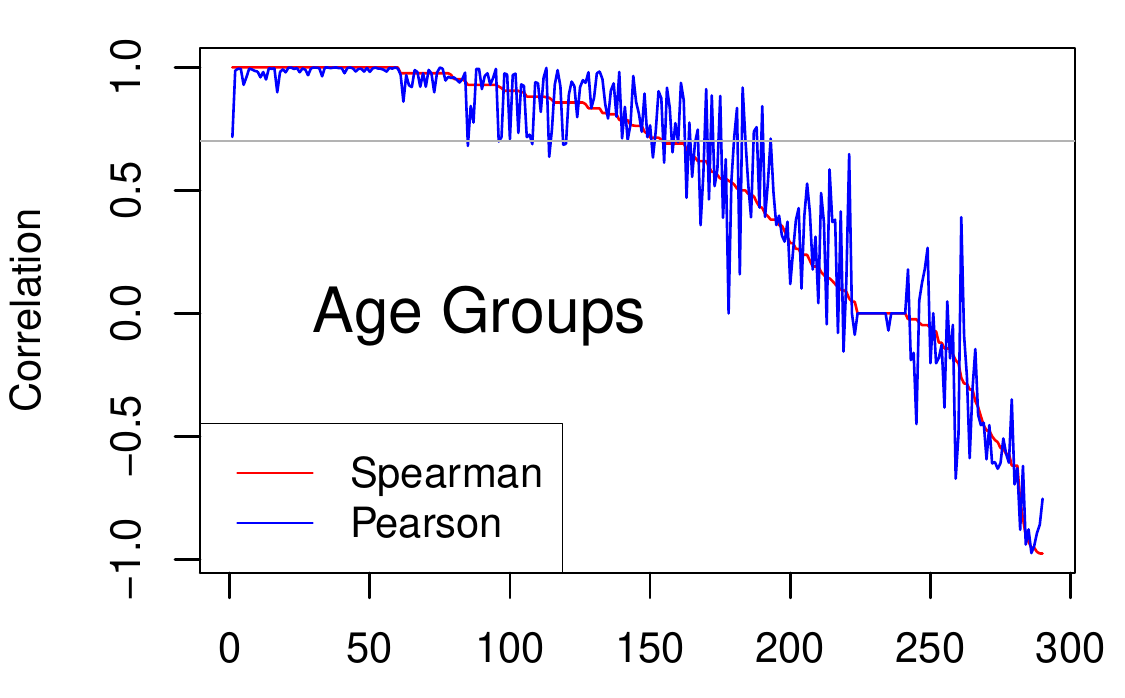}\hspace{0.2cm}
\includegraphics[width=0.48\linewidth]{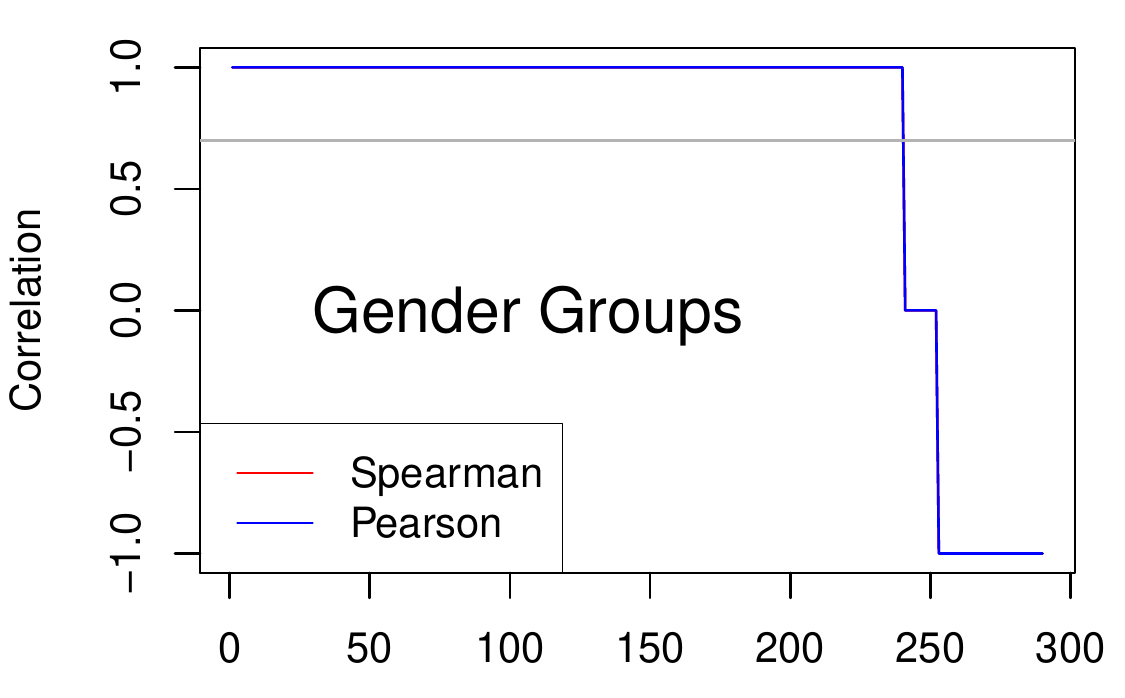}
\caption{Spearman rank and Pearson correlations of normalized interest-specific audiences between two data samples, differentiating among age groups or genders. Grey line denotes 0.70 Spearman correlation level.}
\label{figure:timecorrs}
\end{centering}\end{figure}

Figure~\ref{figure:timecorrs} shows the Pearson and Spearman correlations between the (normalized) audience sizes of the two samples, sorted by correlation magnitude. On the left the correlation is computed between the eight age groups, and right the two genders (keeping all other variables constant). In the comparison of the age groups, we find 45\% of the comparisons to be below 0.70 mark (point at which $p$ gets under 0.05 level), and 17\% even negative, meaning the ordering of the age groups flipped. The gender, being a binary variable, can only be observed to retain or switch their ordering, and for 13\% of cases the latter happened. 

Moreover, to understand the reasons of such differences between the two samples we asked the following question: Are the interests with lower audiences more volatile? In order to respond this question, we correlated the mean of the Spearman correlations given age group for each mark interest and their average audience among all countries. As result, we came with a negative Pearson correlation of -0.57 (p-value 0.001). Thus, surprisingly, interests with higher audience are more unstable.

Intuitively, if Facebook changes its black box often (for example, redefining the interests), then wild fluctuations may happen. But if the changes reflect the actual changes in user base and their evolving interests, then the changes should be similar in places with similar culture, geography, and language. For example, Google Trends shows dieting spikes at the beginning of each year\footnote{https://trends.google.com/trends/explore?q=diet}.

Hence, we compare the changes in interest from one sample to another between United States and United Kingdom (two Western countries sharing a language), and between US and India and Brazil. To do this, we calculated the correlation between the audience variation (the deltas) of one country to another country for each interest. For US vs.\ UK comparison, out of 29 interests, we found 17 to be significantly similar (having Pearson $r>0$ at $p<0.05$) and only one interest to be significantly negatively related between the two countries, implying a similarity in the \emph{direction of change} over time. For US vs.\ India, however, only 5 interests shared a direction (1 negatively), and US vs.\ Brazil 9 (3 negatively). This suggests that comparisons between more similar countries may be more stable in the same time span.


Thus, upon a closer inspection we find more stability in the data across demographics \emph{without including interests}. This may be due to the definition of such features to be stable in time, and possibly to Facebook's inference algorithms for age and gender being relying on more explicit user-specified information. However once we include interest information, the stability of the audience estimates is much worse, with only 45\% of the interest-specific age groups having the same ordering (Spearman's $\rho > 0.7$).

\section{Concluding Discussion}

This paper explores the use of Facebook ad audience estimates for global lifestyle health concern surveillance. Particularly, we assess the quality of Facebook data by introducing placebo baselines, normalization alternatives, and performing temporal analysis. Among our findings, we show that within-country statistics are more statistically separable than statistics across countries, which is an important observation as previous efforts have showed very promising results, but they focused only on a single country analysis~\cite{gittelmanetal15jmir,chunaraetal13pone}. More important, the high volatility observed between two data snapshots warrants extra caution in the use of Facebook Marketing API as a source of social interest.

We can only speculate about the causes of such instability. It is possible that Facebook is constantly updating and changing its NLP pipeline, unevenly improving its performance over different topics. Further, if the time span implied in the interest estimate is short-term, audience numbers may swing wildly as Facebook usage changes, say, during holiday seasons. Another source of variability could come from the redefinition of what content is included in an interest. For instance, if new brands are included in plus-size clothing interest, its volume will increase. Finally, the advertising market demand may drive Facebook to focus on developing one interest over another. 

Still, it may be too early to give up on this potentially rich data source. Our effort is restricted to a specific kind of disease surveillance. Thus, we acknowledge that it is impossible to generalize the limitations we have found without future studies. For example, much as the search volume revealed by Google Trends \cite{carriere2013nowcasting}, the sways in Facebook's interests may be seasonal or local. Unfortunately currently neither Facebook's Advertising interface or Marketing API provide historical data, so more work needs to be done to monitor the data -- i.e.\ for over a year if one is to capture seasonal variation. Furthermore, to assess the accuracy of interest assignment, one could use the advertising platform to run a survey assessing the interests of the reached audience directly.

Thus, we hope that this work will encourage future efforts to use our methodology to gather user interest from the Facebook Ads for other applications and scenarios. We also hope that results presented here will encourage future researchers to test the reliability of this potentially rich data source, and inspire more rigor in the future data-driven studies that aim at correlating social media data with offline world data.

\bibliographystyle{ACM-Reference-Format}
\bibliography{fbglobalhealth} 
\end{document}